\documentstyle[aps,prl,preprint]{revtex}

\begin{document}
%@twocolumnfalse\endcsname

\title{Primordial Magnetic Fields, Right Electrons, and the Abelian
Anomaly}

\author{M. Joyce$^{1,2}$ and M. Shaposhnikov$^1$}

\address{$^1$Theory Division, CERN, 1211 Geneve 23, Switzerland}
\address{$^2$School of Mathematics, Trinity College, Dublin 2,
Ireland}
\date{February 1997}
\maketitle
\begin{abstract}
In the standard model there are charges with abelian anomaly only
(e.g. right-handed electron number) which are effectively conserved
in the early universe until some time shortly before the electroweak
scale. A state at finite chemical potential of such a charge,
possibly arising due to asymmetries produced at the GUT scale, is
unstable to the generation of hypercharge magnetic field. 
Quite large magnetic fields ($\sim 10^{22}$ gauss at $T\sim 100$
GeV with typical inhomogeneity scale $ \sim \frac{ 10^6}{T}$)
can be generated. These fields may be of cosmological interest,
potentially acting as seeds for amplification to larger scale
magnetic fields through non-linear mechanisms. Previously 
derived bounds on exotic $B-L$ violating operators 
may also be evaded.
\end{abstract}
%\vskip0.5pc]

\vspace*{-17.5cm}
\hspace*{9cm} 
\mbox{CERN-TH/97-31, astro-ph/9703005}
\newpage
\noindent
It is usually assumed that the early Universe at temperatures above
the electroweak scale and below, say, $10^{12} -10^{16}$ GeV
(depending on the model of inflation) consists of 
an (almost) equilibrium primordial plasma of elementary particles, 
in which any long-range fields are absent. 
One exception is in the context of the problem of
generating galactic magnetic fields, which may require 
the presence of primordial seed magnetic fields which 
are subsequently amplified by a galactic dynamo mechanism (see, e.g.,
\cite{zel}). The creation of long range magnetic fields requires
that conformal invariance be broken 
in the coupling of the electromagnetic field to gravity \cite{turner}, 
and a number of mechanisms based on different ideas about this 
breaking have been proposed to date \cite{turner,ratra}.
In this letter we argue that there may be a relation
between the appearance of magnetic fields in the early
Universe and two other, apparently completely unrelated, phenomena : (i)
The smallness of the electron Yukawa coupling constant, and (ii) possible
lepton asymmetry of the early Universe.

In short, the logic goes as follows. There are three exact conservation
laws in the standard electroweak theory. The associated conserved charges
can be written as $N_i=L_i - \frac{1}{3} B$, where $L_i$ is the 
lepton number of $i$th generation and B is the baryon number. 
The fourth possible combination,
$B+\sum_i L_i$ is not conserved because of electroweak anomalous
processes, which are in thermal equilibrium in the range $100$ GeV
$<T<10^{12}$ GeV \cite{krs}. Now, if $h_e =0$, where $h_e$ is the right
electron Yukawa coupling constant, then the electroweak theory on the
classical level shows up a higher symmetry, associated with the chiral
rotation of the right electron field. For the small actual 
value of the Yukawa
coupling ($h_e = 2.94 \times 10^{-6}$ in the MSM) this symmetry has an
approximate character. At temperatures higher than $T_R \simeq 80$ TeV
perturbative processes with right electron chirality flip are
slower than the expansion rate of the Universe \cite{cdeo}, and
therefore this symmetry may be considered as an exact one on the 
classical level at $T>T_R$ \cite{mj}. 
(The importance of this symmetry 
for the consideration of the wash-out of the GUT baryon asymmetry by 
anomalous electroweak $B$ and $L$ non-conserving reactions was 
realized in refs. \cite{iban,cdeo,cko}.) Suppose now that an excess 
of right electrons over positrons was created by some means at $T>T_R$ 
( e.g. by a GUT mechanism for baryogenesis).
Now the right electron number current $j^{\mu}_R$ is violated in the minimal 
standard model (MSM) as described by the anomaly equation 
\begin{equation}
\partial_\mu j^\mu_R = -\frac{g'^2 y_R^2}{64 \pi^2} f_{\mu
\nu} \tilde{f}^{\mu \nu},
\end{equation}
where $f$($\tilde{f}$) are the $U_Y(1)$ hypercharge field strengths
(and their duals) respectively, $g'$ is the associated gauge
coupling and $y_R=-2$ is the hypercharge of the right electron. 
The number of the right electrons $N_R$  therefore changes
with the Chern-Simons (CS) number of the hypercharge field configuration
as $\Delta N_R = \frac{1}{2} y_R^2\Delta N_{cs}$
with
\begin{equation}
N_{cs}=-\frac{g'^2}{32 \pi^2}\int d^3 \vec{x} \epsilon_{ijk}
f_{ij}b_k,
\label{integratedanomaly}
\end{equation}
where $b_k$ is the hypercharge field potential. 

One can now see qualitatively that there is an instability in hot matter 
with an excess of right electrons towards formation of hypercharge fields 
with CS number as follows. %we will discuss. 
(The line of reasoning presented here is similar to
the consideration of cold fermionic matter with anomalous charges in
\cite{RubandTav}). The energy density ``sitting" in 
right electrons with a chemical potential $\mu_R$ is of order 
$\mu_R^2T^2$, and their number density of order $\mu_R T^2$. 
On the other hand this fermionic number can be absorbed by a 
hypercharge field of order $g'^2 k b^2$, with energy of order
$k^2 b^2$, where $k$ is the momentum of the classical hypercharge
field and $b$ is its amplitude. Therefore, at $ b> T/g'^2$ and 
$k \sim \mu_R T^2/(g'^2 b^2)$ the gauge field configuration has the same
fermion number as the initial one, but smaller energy. An
instability to 
generation of hypercharge magnetic field, which tends to ``eat up"
real fermions, results.  It is important here that at temperatures
$T>T_R$ the electroweak symmetry is ``restored", and that the U(1)
hypercharge magnetic field is massless at that time. (No term like
$m_Y^2 b^2$ is generated in any order of perturbation theory in
abelian gauge theory at high temperature \cite{fradk}; the lattice
study in \cite{u1} confirmed this expectation for SU(2)$\times$U(1)
EW theory beyond perturbation theory). If the hypercharge magnetic
fields survive until the time of the EW phase transition ($T\sim 100$
GeV), they will give rise to ordinary magnetic fields because of
electroweak mixing. In the rest of this paper we present quantitative
estimates of the (hypercharge) magnetic fields which may be produced
by this effect.

Let us discuss first the possible origin and the magnitude of the 
the required right electron number asymmetry $\delta_R=e_R/s$, where
$s=\frac{2}{45}\pi^2 T^3 N_{eff}$ is the entropy density with 
$N_{eff}=N_b + \frac{7}{8}N_f= 106.75$ the total effective number
of degrees of freedom of the MSM. In principle $\delta_R$ produced by
out of equilibrium decay at the GUT scale can be as large as $\sim
10^{-2}-10^{-4}$ (for a review see, e.g. \cite{gutbaryo}). This is
quite consistent with the magnitude of the final baryon asymmetry
$\delta_B$ being that observed since there is no simple general
relation between the two numbers. In theories like those discussed in
\cite{cdeo,cko} with $L$ violating processes at intermediate scales
one has $\delta_B \sim \delta_R$, at least in the case that the $L$
violating processes go out of equilibrium before the $e_R$ violating
ones come into equilibrium. In \cite{cko} the case is considered
where the L violation continues for just long enough to reduce the
final $\delta_B$ to the observed one from an initially larger value
fixed by $\delta_R$. And, in a simple GUT like $SU(5)$ in which 
the charges $N_i=L_i-\frac{1}{3}B$ are conserved, we can have 
$\delta_B=0$ at the electroweak scale irrespective of the value 
of $\delta_R$ during the time it is effectively conserved.
In the rest of this letter we will simply assume the 
existence of a primordial density of $e_R$, with its chemical 
potential as a free input parameter, assumed
only small enough to be treated perturbatively. We also assume that no
hypercharge magnetic fields existed before the right electron excess is
generated.

The effect of the anomaly on the gauge field dynamics is given 
through the term in the effective Lagrangian 
\begin{equation}
{\delta\cal{L}} = \frac{g'^2}{4 \pi^2} \mu_R
\epsilon_{ijk} f_{ij} b_k .
\label{redl}
\end{equation}
which is obtained by integrating out the
fermions at finite chemical potential \cite{Redlich}.
It simply describes how winding the gauge fields to give CS number
changes the energy of the system because it changes the number of
fermions as described by the anomaly equation. Adding this term to
the ordinary Lagrangian for the gauge fields leads to the equations
of motion 
\begin{eqnarray}
\frac{\partial\vec{H}}{\partial t} + \vec{\nabla} \times \vec{E}=0,
\label{max} \\
\vec{E}=\frac{1}{\sigma}(\vec{\nabla} \times \vec{H} + 
\frac{4 \alpha'}{\pi}\mu_R \vec{H}).
\label{electric}
\end{eqnarray}
where $\alpha'={g'^2}/{4 \pi}$. These are simply Maxwell's equations
with the additional term due to the anomaly ($\propto \mu_R \vec{H}$) and
the assumption that the total (hypercharge) current is given by 
$\vec{j}=\sigma \vec{E}$, where $\sigma$ is the conductivity of
the plasma and $\vec{E}$ is the (hyper-)electric field. 
We have also dropped the term $\partial\vec{E}/\partial t$
since (as we will see below) the fields always evolve on a time-scale
which is much longer than $\sigma^{-1}$. In the expanding 
FRW Universe with scale factor $a$
the equations have exactly the same form in conformal time coordinates 
$\tau= \int a^{-1}(t) dt$, but with the replacements 
$\mu_R \rightarrow \mu_Ra$  and $\sigma (\propto T) \rightarrow \sigma a$.
The fields $\vec{E}$ and $\vec{H}$ are those given by their 
standard definitions in the conformal frame which will be 
related to the physical fields at the appropriate point below. 
We also have the following kinetic equation for $\mu_R$ : 
\begin{equation}
\frac{1}{a}\frac{\partial(\mu_R a)}{\partial \tau} = -\frac{\alpha'}{\pi}\frac{783}{88}\frac{1}{a^3 T^2} 
\vec{E}\cdot\vec{H} - \Gamma_R (\mu_R a) ,
\label{elnumber}
\end{equation}
in which the first term describes the change in the chemical potential
due to the anomaly ($f \tilde{f} \propto \vec{E}\cdot\vec{H}$), and the 
second the change due to the perturbative processes
which flip electron chirality with the rate $\Gamma_R=\frac{T_R}{M_0}T$
($M_0= M_{pl}/1.66 \sqrt{N_{eff}}\simeq 7.1 \times 10^{17}$ GeV). The
numerical coefficient $\frac{783}{88}$ comes from the relationship 
between right electron chemical potential and right electron number
asymmetry (in terms of which the anomaly is expressed)
\[
\mu_R=\frac{2}{45}\pi^2 N_{eff}[\frac{783}{88}\delta_R -
\frac{201}{88}\delta_1 + \frac{15}{22}(\delta_2+\delta_3)] T,
\]
which is obtained from a local thermal equilibrium calculation in the EW
theory with three fermionic generations and one scalar doublet, with
the conserved charges assumed to be $N_i(=\delta_i s)$ and $e_R(=\delta_R s)$.

With a Fourier mode decomposition 
$\vec{H}(\vec{x})= 
\int d^3\vec{k} \vec{H}(\vec{k}) e^{-i\vec{k}.\vec{x}}$ with
$\vec{H}(\vec{k})=h_i \vec{e}_i$ where $i=1,2, \vec{e_i}^2=1,
\vec{e}_i\cdot \vec{k}=0, \vec{e}_1\cdot\vec{e}_2 =0$, 
the linear equations (\ref{max}) and (\ref{electric}) become   
\begin{eqnarray}
\partial_\tau h_1 + \frac{k^2}{\sigma a}h_1 - 
\frac{4 i \mu |k|}{\sigma}h_2=0
\nonumber
\\
\partial_\tau h_2 + \frac{k^2}{\sigma a}h_2 + 
\frac{4 i \mu  |k|}{\sigma}h_1=0
\label{equat}
\end{eqnarray}
where $\mu\equiv \frac{g'^2}{4 \pi^2} \mu_R$. 
The mode
\begin{equation}
h_2(\tau,k) = -i h_1(\tau,k)=-\frac{i}{2}(h_1(0,k) + i h_2(0,k)) 
exp(\lambda_+ (\tau))
\label{modes}
\end{equation}
where
\begin{equation}
\lambda_{\pm}(\tau)=-\frac{k}{\sigma a}(k \tau \mp 4\int_0^\tau d\tau' \mu a)
\label{lambda-pm}
\end{equation}
is an unstable mode which is growing at conformal time $\tau$ if
$k < 4 \mu (\tau) a (\tau) $. 
%, growing with exponent $\xi_+\tau$. 
It has the property 
$\vec{E}(\vec{k})= \frac{1}{\sigma a}(-|k|+4\mu a) \vec{H}(\vec{k})$. 
(The other orthogonal mode decays at any $|k|$.) 

Consider now the approximation in which the chemical potential
$\mu$ is a constant. The growing instability starts to
develop at $T \sim T_g$ where we define $T_g$ to be
\begin{equation}
8(\frac{\mu}{T})^2
	\frac{1}{\sigma/T} \frac{M_o}{T_g} = 1.
\label{begins}
\end{equation}
(when the maximally growing mode with $k=2\mu a$ has begun growing
significantly).
A necessary requirement for the instability to develop is that 
$T_g > T_R$, since if this is not satisfied the second term in
(\ref{elnumber}) will rapidly reduce $\mu$ towards zero. 
Translated into a minimum value for $\delta\equiv \frac{\mu}{T}$ this 
requires $\delta > \delta_{crit}= 10^{-6}$ (using
$\sigma \approx 68 T$ \cite{jpt}). For $\delta < \delta_{crit}$
no non-trivial dynamics result from the presence of such a chemical
potential since the unstable modes are frozen on the relevant time-scale.
If $\delta > \delta_{crit}$ the evolution of the instability 
for $T < T_g$ will be given by the simple growth factor above,
until the time at which the growth becomes significant enough
that the first term in (\ref{elnumber}) is important. To estimate 
when this is and what the amplitude of the field is at that time
it is sufficient to calculate the CS number as a function  of time.
It is given (per co-moving volume) by 
\begin{eqnarray}
 n_{cs}(\tau) = -\frac{g'^2}{ 32  \pi^2}
<\epsilon_{ijk} f_{jk}(\tau) b_i (\tau)>
\nonumber
\\
\approx -\frac{g'^2}{64\pi^4} \int_0 ^{4\mu a}
 dk e^{2 \frac{k(4\mu a - k) \tau }{\sigma a}}k^2 f(k),
\label{csnumber}
\end{eqnarray}
neglecting all but the growing mode. We have also taken
$<b_i(\vec{k}, \tau ) b_j^*(\vec{l}, \tau )>|_{\tau=0} =
\delta^3(\vec{k} - \vec{l}) \delta_{ij} <b^2(k)>_0$, assuming
translational and rotational invariance of the initial perturbations,
and assumed that the perturbations are thermal in origin, with the
appropriate normalizations, $<b^2(k)>_0= \frac{1}{2k (2 \pi)^3} f (k)$
where $f(k)= (e^{\frac{k}{T_0}}-1)^ {-1}$ is the bosonic distribution
function and $T_0$ the temperature at which we define $a_0=1$. Defining
$\epsilon$ by $\frac{1}{2} y_R^2 n_{cs} \equiv \epsilon \Delta e_R
a^3$, where $\Delta e_R = \frac{88}{783} \frac{\mu_R}{T} T^3$ i.e. the
difference between the right electron density in the initial state and
the $\mu_R = 0$ state, the linear approximation breaks down when
$\epsilon \sim 1$. Evaluating the integral in (\ref{csnumber}) we find
$\epsilon \approx 2 \times 10^{-6} \delta
\frac{1}{\sqrt{\alpha}}e^\alpha$ where $\alpha=T_g/T$ with $T_g$ as in
(\ref{begins}). Thus for a few expansion times after the mode
starts growing at temperature $T_g$ we have $\epsilon < 1$ and the
linear approximation is valid. The corresponding physical magnetic
field $H_{phy}$ can be estimated by putting $|n_{cs}|\approx
\frac{g'^2}{16 \pi^2} k b^2 = \frac{1}{2} \epsilon \Delta e_R a^3$ and
using $kb=a^2 H_{phy}$, where $k \sim 2\mu a$ (i.e. assuming the
maximal growing mode to dominate). Putting in the numbers this gives a
physical magnetic field of strength $H_{phy} \approx 2 \times 10^2
\sqrt{\epsilon \delta \frac{k_{phy}}{T}} T^2$ at a physical length scale
$k_{phy}^{-1} \equiv(\frac{k}{a})^{-1} 
\sim \frac{1}{2\delta T}$. For $\delta \sim 5\times
10^{-6}$ we are in the linear regime until $T_R \sim 80$ TeV and at
that point therefore have a magnetic field of strength $H_{phy} \sim
6 \times 10^{26}$ gauss ($1$ GeV$^2$ $=1.95\times 10^{20}$ gauss) at a length
scale of $\sim 10^{5}/T$ (compared to a horizon scale of
$\sim 10^{13}/T$). 

How do the fields evolve for $T < T_R$? In the case that  
the linear ( i.e. constant chemical potential) approximation
is good, one expects that the growth will rapidly turn into 
decay as $\mu$ is damped. Within a few expansion times the growth
will be undone as the maximally growing mode $k \sim 2\mu a$
now decays with exponent $-(k^2/\sigma) a \tau$. What about the
case when this linear approximation breaks down? 
To treat this case we must analyse the full non-linear set of equations 
(\ref{max})-(\ref{elnumber}). We have done this numerically
with the simplifying assumption that  
the distribution of right electron number is homogeneous in space. 
Then the two linear equations (\ref{max}) and (\ref{electric}) can be 
solved exactly for any time dependent $\mu$, the solution inserted 
in (\ref{elnumber}), and the averaging over thermal initial conditions 
performed. The resulting equation is
\begin{equation}
\frac{\partial\mu a}{\partial \tau} = 
(\frac{\alpha'}{\pi})^2 \frac{783}{88}\frac{1}{32\pi^2\sigma a^2 T}
\int_0^\infty dk k^2[(k - 4\mu a)exp(2\lambda_+(\tau))-
(k + 4\mu a)exp(2\lambda_-(\tau))],
\end{equation}
a length
scale of $10^{5}/T$a length
scale of $10^{5}/T$with $\lambda_{\pm}(\tau)$ given by (\ref{lambda-pm}).
Our results show that the chemical 
potential, typical physical momentum of the magnetic field configuration 
and the magnetic field energy  scale as
\begin{equation}
\frac{\mu_R}{T} \propto \frac{k_{phy}}{T}\propto \frac{H_{phy}^2}{T^4}\propto \left(\frac{T}{T_g}\right)^{\frac{1}{2}}.
\end{equation}
in the range  $T_g > T > T_R$. This behaviour can be easily understood
qualitatively as follows. As the instability develops, the linear 
approximation breaks down and $\mu$ starts significantly decreasing.
This shifts the growth of modes to longer wavelengths. This procedure
continues, growth of any mode eventually turning itself off
and increasing the growth coefficient of modes at larger scales.
The minimum value of $\mu$ which can be reached at any given time 
$\tau$ (and, correspondingly, the maximum physical scale for the 
sourced fields) is simply that given by (\ref{begins}), solved for 
$\mu$ with $T_g$ replaced by the temperature $T(\tau)$ i.e it is 
just the minimal chemical potential required to drive a growing mode 
at that time in the linear approximation. The parametric dependence on the
temperature observed follows from the fact that the chemical potential 
(and maximally growing mode) trace these values. The dependence of the
magnetic field energy follows from the expression we derived in
the previous paragraph by setting the CS number of the configuration 
to cancel the total fermion number, but now taking the appropriate
scaling for $k_{phy}$ itself.

Evolving the system forward from $T_R$ to the electroweak scale
$T_{ew}$ we see the damping of the fields for 
$\delta$ in the linear regime anticipated above, as the perturbative 
processes erase the chemical potential driving the growth.
As $\delta$ increases, however, this damping becomes less
efficient, and for $\delta > 2\times 10^{-4}$ we find that the 
damping has not set in at all by the electroweak scale. 
The reason for this behaviour is also simple. For a mode which evolves
in the linear regime, the growth and decay exponents are effectively
the same for the modes which grow significantly. Once we enter the
non-linear regime this is no longer true, since the maximally growing
mode carries the integrated effect of growth until any given time
i.e. it has grown with exponent ${k}\int \frac{\mu(\tau)}{\sigma} d\tau$
which is much greater than  $\frac{k^2}{\sigma a} \tau$. 
Put another way, the mode has
been able to grow on a scale significantly larger than the diffusion
length for magnetic field at the relevant time, and it takes some time
after the end of the growth for the latter scale to catch up and undo
the effect of the instability.
For $\delta$ in this region we also see that the typical scale of 
magnetic fields $k$ and value of the chemical potential at $T_{ew}$ 
do not depend on initial asymmetry, 
$\frac{2k_{phy}}{T}\simeq \frac{4\mu}{T} \simeq\frac{10^6}{T}$. 
The amplitude of magnetic field scales as $H \propto \sqrt{\delta}$ and
e.g. for $\delta = 10^{-2}$ we find $ H \simeq 4\times 10^{22}$ gauss.
This is as we would expect from the discussion of the scaling above.

What is the ultimate fate of these magnetic fields? 
Unless some other effect comes into play in the dynamics, the fields 
will decay. One such effect is turbulence. With the full set of 
MHD equations (which include the velocity of the fluid which we 
have neglected) there is a transition to a turbulent regime 
when the magnetic Reynold's number $R= \sigma L v$ is large \cite{ol1}.
The reason we have evolved the equations to the electroweak scale is
that, if the electroweak phase transition is of first order, it serves 
as a source of turbulence \cite{bbmcewturb}. Since we have here 
$\sigma \sim 10^{2} T$ and magnetic fields which 
begin to grow on length scales $L$
up to $\sim 2 \times 10^{6}/T$ we expect to enter the turbulent regime
if there are bulk velocities of greater than $\sim 5\times 10^{-9}$,
which are certainly larger that the expected velocity of the bubble
walls. A recent study of this phenomenon \cite{ol1} suggests that the
effect of this turbulence is to transfer the magnetic energy to larger
length scales, thus evading the Silk argument \cite{silk}. If true, the
fields generated by the mechanism under discussion may play the  role of
the seed galactic magnetic fields. Note that the seed fields we obtain at
the electroweak scale with the mechanism we have discussed 
($\sim 10^{22}$ gauss) are much larger than those
generated at bubbles walls ($\sim 10^{-2}$ gauss) which were suggested 
as seeds for amplification through turbulence in \cite{bbmcewturb}.
It is also worth mentioning the particular structure of the magnetic fields 
appearing because of the abelian anomaly. The CS wave (\ref{modes}) has a
non-zero value of $\vec{H}\cdot \vec{\partial}\times\vec{H}$ and thus 
breaks parity. Could it be that the rotation of galaxies are related to this?
Study of the entire set of MHD equations with the 
additional anomalous terms discussed in this letter will be required
to address this question.

Finally let us mention that the processes we have considered 
also affect the bounds on the strength of exotic interactions with
$B-L$ violation derived from the requirement that GUT baryon asymmetry
is not erased by sphalerons \cite{cdeo,cko} (which is important if no
baryon asymmetry is created at the electroweak scale). If the
right electron asymmetry produced at the GUT scale is small enough
($\delta < 10^{-6}$), then the bounds are obviously not affected since
abelian anomaly does not play any role. If, on the other hand, $\delta$
is large enough that significant CS number survives remains in
the condensate until the electroweak scale,
any bound on the strength of exotic interactions can be evaded.
Irrespective of the effect of any $B-L$ violation until that point
the remaining CS number will be converted into quarks and leptons 
carrying net baryon number at the electroweak phase transition.
The final baryon asymmetry will depend on the initial value of
$\delta$ and the exact strength of the phase transition (which will
determine how the $B$ violating processes turn off). Conversely, given
detailed knowledge of the phase transition, it will be possible to
place an upper bound on the initial value of $\delta$ in the very
early universe, and on the strength of the magnetic fields resulting
at the electroweak phase transition. 

We are grateful to K. Enkvist, M. Giovannini, A. Kusenko and L. McLerran for
interesting discussions.


\begin{thebibliography}{99}
\bibitem{zel} S.I. Vainshtein and Ya. B. Zeldovich, Usp. Fiz. Nauk
{\bf 106}:431, 1972

\bibitem{turner} M.S. Turner and L.M. Widrow, Phys. Rev.
{\bf D37}:2743, 1988

\bibitem{ratra} B. Ratra, The Astronomical J., {\bf 391}:L1, 1992;
A. Dolgov and J. Silk, Phys. Rev. {\bf D47}:3144, 1993;
M. Gasperini, M. Giovannini and G. Veneziano, Phys.
Rev. Lett. {\bf 75}:3796, 1995;
D. Lemoine and M. Lemoine, Phys. Rev. {\bf D52}:1955,
1995

\bibitem{krs}
V.~A. Kuzmin, V.~A. Rubakov and M.~E. Shaposhnikov,
Phys. Lett., {\bf 155B}:36, 1985

\bibitem{cdeo}{B. Campbell, S. Davidson, J. Ellis and K. Olive, Phys.
Lett. {\bf 297B }:118, 1992}

\bibitem{mj}{In alternative cosmologies such as that considered in 
M. Joyce, Phys. Rev. {\bf D55}:1875, 1997,  in which
the expansion rate before nucleosynthesis is different to the standard
one, $T_R$ can be well below the electroweak scale.}

\bibitem{iban}
L.~E. Ibanez and F.~Quevedo,
Phys. Lett., {\bf B283}:261, 1992

\bibitem{cko}{J. Cline, K. Kainulainen and K. Olive, Phys. Rev. Lett.
{\bf 71}:2372, 1993; Phys. Rev. {\bf D49}:6394, 1994}

\bibitem{RubandTav}V. Rubakov and A. Tavkhelidze, Phys. Lett.  {\bf
165B}:109, 1985;
V. Rubakov, Prog. Theor. Phys. {\bf 75}:366, 1986

\bibitem{fradk}E. S. Fradkin, Proc. Lebedev Inst. {\bf 29}:1, 1965

\bibitem{u1} K. Kajantie, M. Laine, K. Rummukainen and M.
Shaposhnikov, Nucl. Phys. {\bf B493}:413, 1997

\bibitem{gutbaryo}E.W. Kolb and M.S. Turner, The Early Universe,
Addison-Wesley, Reading, MA, 1990.

\bibitem{Redlich}{A.N. Redlich and L.C.R. Wijewardhana, Phys. Rev.
Lett. {\bf 54}:970, 1985}

\bibitem{jpt}{M. Joyce, T. Prokopec and N. Turok, Phys. Rev.
{\bf D53}:2930, 1996; G. Baym and H. Heiselberg, astro-ph/9704214}

\bibitem{bbmcewturb}{G. Baym, D. Bodecker and L. McLerran,
Phys.Rev. {\bf D53}:662, 1996}

\bibitem{ol1} P. Olesen, astro-ph/9610154;
 A. Brandenburg, K. Enqvist and P. Olesen,
hep-ph/9608422; Phys. Rev. {\bf D54}:1291, 1996

\bibitem{silk} J. Silk, Astrophys. J. {\bf 151}: 459, 1968

\bibitem{ht} J. Harvey and M. Turner, Phys. Rev. {\bf D 42}: 3344 (1990);
B. Campbell, S. Davidson, J. Ellis and K. Olive, Phys.
Lett. {\bf 256B }:457, 1991; W. Fischler, G.F. Giudice, R.G. Leigh, and S. Raban, Phys. Lett. {\bf B258}: 45 (1991) 

\end{thebibliography}
\end{document}